\documentclass[aps,prl,twocolumn,groupedaddress,notitlepage,nofootinbib,showpacs]{revtex4-1}
\usepackage{graphicx,epsf,color,amsmath,multirow}
\usepackage[caption=false]{subfig}
\usepackage[yyyymmdd,hhmmss]{datetime}
\newif\ifdraft
\drafttrue
\newif\ifpreprint
\preprinttrue

\def\Section#1{\noindent\textsl{#1.\/}}

\def\fig#1{Fig.~{\ref{#1}}}

\def\spa#1.#2{\left\langle#1\,#2\right\rangle}
\def\spb#1.#2{\left[#1\,#2\right]}
\def\tree{{\rm tree}}
\def\oneloop{{\rm 1\hbox{-}loop}}

\def\NP{{\rm NP}}
\def\P{{\rm P}}
\def\NYM{{\cal N}_{\rm YM}}
\def\Ord{\mathcal{O}}

\def\eqn#1{Eq.~(\ref{#1})}

\def\eqns#1#2{Eqs.~(\ref{#1}) and~(\ref{#2})}

\def\NeqFour{{{\cal N}=4}}

\def\eps{\epsilon}
\def\nn{\nonumber}
\def\Section#1{\vskip .2 cm 
\noindent{\em #1:}}

\def\spa#1.#2{\left\langle#1\,#2\right\rangle}
\def\spb#1.#2{\left[#1\,#2\right]}
\def\deltaQ{\delta^{(8)} (Q)}

\def\oneloop{(1)}

\def\pYM{\rm YM}
\def\NYM{\pYM}

\def\NP{\rm NP}
\def\P{\rm P}

\def\tb{{\bar t}}

\def\ct{{\rm ct}}
\def\NMHV{N${}^k$MHV}

\def\eps{\epsilon}
\def\Li{\mathop{\textrm{Li}}\nolimits}

\begin{document}

%\title{anomalous amplitudes in ${\cal n}=4$ supergravity beyond one loop}
\title{Two-loop $n$-point anomalous amplitudes in $\mathcal{N}=4$ supergravity}

\author{Zvi~Bern${}^{a }$, David Kosower${}^{b}$ and Julio Parra-Martinez${}^{a }$ }
\affiliation{
$\null$\\                                                                                                                   
${}^a$Mani L. Bhaumik Institute for Theoretical Physics,\\
UCLA Department of Physics and Astronomy,\\
Los Angeles, CA 90095, USA\\
\\
${}^b$\hbox{Institut de Physique Th\'eorique, CEA, CNRS, Universit\'e Paris--Saclay},
  F--91191 Gif-sur-Yvette cedex, France \\
}

\begin{abstract}
We compute the anomalous two-loop four-point amplitudes in $\NeqFour$ pure
supergravity, using unitarity and the double-copy construction.  We also
present all terms determined by four-dimensional cuts in two all-multiplicity
two-loop anomalous superamplitudes. This result provides the first two-loop $n$-point
gravity amplitude, up to a class of undetermined rational terms, which are absent
at four points.  We show that a recently proposed finite counterterm cancels
these amplitudes to this order.  We argue that the counterterm does not spoil
the three-loop finiteness of anomalous amplitudes in the $\NeqFour$ theory. 
\end{abstract}


\maketitle

\Section{Introduction}
All but the simplest perturbative calculations in quantum gravity
using standard Lagrangian methods quickly lead into a swamp of
intractable intermediate expressions.  Modern approaches, in
particular generalized unitarity~\cite{BDDK,MoreGeneralizedUnitarity}
and double-copy relations between gravity
and gauge theories~\cite{KLT,BCJ,BCJLoop}, have made possible multi-loop
calculations~\cite{BDDPR}, as well as loop-level results with large
numbers of external legs~\cite{CollinearGravity}.  

Multiloop amplitudes in supergravity theories offer a 
window on ultraviolet properties~\cite{SupergravityN8UV,
  ThreeLoopN4Sugra, Enhanced, FiveLoops}.  Supersymmetry constraints
push possible divergences to higher-loop order; but are they the whole
story?  Extended supergravity theories display nontrivial enhanced cancellations of
ultraviolet singularities~\cite{ThreeLoopN4Sugra, Enhanced}.
These have yet to be understood from standard
symmetry arguments, despite valiant attempts~\cite{UVSymmetry}. (See Ref.~\cite{Kallosh:2018wzz} for recent progress.)  The only
explicit divergence found in $D=4$ pure supergravity is in the
$\NeqFour$ theory~\cite{FourLoopN4Sugra}.  This theory possesses
another important feature, not found in supergravity theories with
more supersymmetries: an anomaly~\cite{MarcusAnomaly} in a $U(1)$
subgroup of its $SU(1,1)$ duality group.  As a consequence, certain
classes of amplitudes that vanish at tree level, thanks to this
symmetry, fail to do so at one loop~\cite{CarrascoAnomaly}. 

As shown in Ref.~\cite{OneLoopAnomaly}, an appropriate finite local
counterterm can cancel the one-loop anomalous amplitudes. This was
shown explicitly for all five-point amplitudes and certain infinite
classes of anomalous amplitudes. The local counterterm appears to
effectively adjust the regularization scheme so as to preserve the
$U(1)$ subgroup responsible for the vanishing of amplitudes.  We may
ask: do the cancellations continue to higher loops?

In this Letter, we evaluate the four-point two-loop anomalous
amplitudes in $\NeqFour$ supergravity, and all but rational
contributions to two all-multiplicity two-loop anomalous amplitudes.
We use the loop version of the Bern-Carrasco-Johansson (BCJ) double
copy~\cite{BCJLoop} to compute the complete four-point amplitudes, and
four-dimensional cuts to compute the polylogarithmic terms of the
$n$-point amplitudes.  These are the first-ever all-multiplicity
results for any gravity amplitudes at two loops.  We show that the
same finite counterterm that removes the anomalous one-loop amplitudes
also removes the two-loop ones.  These results make it plausible that
all anomalous amplitudes are completely removed by the counterterm of
Ref.~\cite{OneLoopAnomaly}.  If this is indeed true, it would have
very interesting consequences for the ultraviolet properties of the
$\NeqFour$ theory, a point to which we return below.

\Section{Review}
The spectrum of pure $\NeqFour$ supergravity~\cite{N4superGrav} has two
supermultiplets:
\begin{equation}
\begin{aligned}
\Phi^+ & = \{\, h^{++}, \psi^{+}_a, A_{ab}^+,\chi^{+}_{abc}, \bar t\, \} \,,\\
\Phi^- & = \{\, h^{--}, \psi^{-}_a, A_{ab}^-,\chi^{-}_{abc}, t \, \} \,,
\end{aligned}
\label{N4multiplets}
\end{equation}
where $\pm$ indicates the helicity;  $a,b,c$ are $SU(4)_R$ symmetry indices; $h$ is the graviton,
$\psi$ the gravitino, $A$ the vector, $\chi$ the spin-$\frac12$
fermion and $(t,\tb)$ the complex scalar.  Using the BCJ
loop-level construction~\cite{BCJLoop}, we can build the $\NeqFour$
theory as a double copy of $\NeqFour$ super-Yang--Mills and pure
Yang--Mills theories~\cite{OneLoopN4, N4Twoloop, ThreeLoopN4Sugra,
  FourLoopN4Sugra}.  The two multiplets in \eqn{N4multiplets}
correspond respectively to the $\NeqFour$ super-Yang--Mills multiplet
tensored with either positive- or negative-helicity gluons.

We may classify amplitude multiplets according to their helicity-violating
(\NMHV) degree, $k =0, \ldots,n-4$ along with the numbers $n_+$ and $n_-$ of
particles in the $\Phi^+$ and $\Phi^-$ multiplets~\cite{CarrascoAnomaly}.   The
former corresponds to the supersymmetric side of the double copy, while $n_\pm$
correspond to the nonsupersymmetric side:
\begin{equation} 
  M^{(n_+,n_-;L)}_{n,k}\equiv
  M_{n,k}^{(L)}(1^-, \ldots, n_-^-, (n_-\!+\!1)^+,
                 \ldots, {n}^+)\,.
\end{equation}
Superscripts denote the type of multiplet, $n = n_+ + n_-$ and $L$ is the loop
order.  In all amplitudes, we omit a factor of $(\kappa/2)^{n+2L-2}c_\Gamma^L/(4\pi)^{L(2-\eps)}$,
where $\eps = (4- D)/2$ is the dimensional regulator and where $c_\Gamma$ is given Eq.~(3.2) of Ref.~\cite{CurvatureSquareN4}. 
In this Letter, we will consider only MHV amplitudes where $k = 0$.
Vanishing amplitudes excluded by supersymmetry Ward
identities are already eliminated by the use of on-shell superspace.
At tree level, the $U(1)$ selection rule~\cite{CarrascoAnomaly} leaves
only the $M^{(n-k-2, k + 2; 0)}_{n,k}$ amplitudes
nonvanishing. The four-point superamplitude at tree level has $k=0$,
\begin{equation}
  M^{(2,2;0)}_{n,0} = i  \frac{\spa1.2^4 \spb1.2}{\spa3.4} \frac{\deltaQ}{\prod_{i<j}\spa i.j}\,.
  \label{eq:treenonanom}
\end{equation}
The spinor inner products $\spa{a}.b$ and $\spb{a}.b$ follow the conventions in
Ref.~\cite{ManganoParke} and $Q$ is the supermomentum in the usual on-shell
superspace~\cite{NairSuperspace}.  Beyond tree level, the anomaly gives
nonvanishing values to other amplitudes.  The tree-level vanishing of
anomalous amplitudes does eliminate all divergences from the corresponding
one-loop amplitudes, which are also purely rational functions of the external
spinors.

Formulas for all but two classes of $k=0$ one-loop superamplitudes
were conjectured based on soft limits in
Ref.~\cite{OneLoopAnomaly}. We will be interested in higher-loop
corrections to the following superamplitudes,
\begin{equation}
\begin{aligned}
  M^{(0,n;1)}_{n,0}&= i (n\!-\!3)! \,\deltaQ\,,\\
  M^{(1,n\!-\!1;1)}_{n,0}&=
%  \\& 
  -i (n\!-\!4)!\,\deltaQ 
  %\\&\hphantom{= -i (n\!-\!4)!\,}\times
  \sum_{r=1}^{n-1} 
  \frac{\hspace{.25pt}\spb n.r \spa r.x \spa r.y}{\spa n.r\!\spa n.x\! \spa n.y}\,. \\
\end{aligned}
\label{eq:oneloopanom}
\end{equation}
The latter expression is independent of the (arbitrary) reference spinors $x,y$.

%%%%%%%%%%%% FIGURE %%%%%%%%%%%%%%
\begin{figure}[tb]
  \centering
%\vskip -.3cm
%  \includegraphics[scale=0.7]{figures/spanning}
  \begin{minipage}{0.42\columnwidth}
  \includegraphics[scale=0.65]{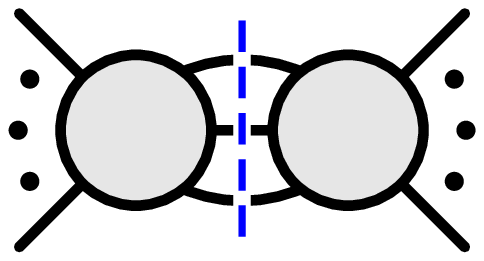}\\ [-.2 cm] (a)
  \end{minipage}
%  \hfill
\hskip .7 cm 
  \begin{minipage}{0.42\columnwidth}
  \includegraphics[scale=0.65]{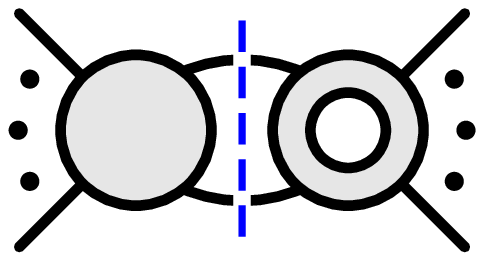}\\ [-.2 cm] (b)
  \end{minipage}
  \caption{Spanning sets of unitarity cuts at two loops: (a) three-particle cuts
           (b) two-particle cuts. The shaded blobs denote tree-level amplitudes,
           and the annulus denotes a one-loop amplitude.}
  \label{fig:spanning}
\end{figure}
%%%%%%%%%%%%%%%%%%%%%%%%%%%%%%%%%%

\Section{Two-loop amplitudes}
We compute the two-loop four-point anomalous amplitude of $\NeqFour$
supergravity using the double-copy construction applied to the amplitudes of
Refs.~\cite{BRY,TwoLoopQCD}, as described in Ref.~\cite{N4Twoloop}. The general
form for the two-loop superamplitudes was presented there,
\begin{equation}
\begin{aligned}
& M^{(n_+, n_-; 2)} (1,2,3,4)
= 
-i s_{12}s_{23}  \,
A_{\NeqFour}^{\rm tree}(1,2,3,4) 
\\& \hskip 5mm \times 
\bigl[
\bigl( s_{12} \, A^{\P}_{1234, \,\NYM} +s_{34}\, A^{\P}_{3421, \,\NYM}\bigr)
\\& \hskip 5mm \hphantom{\times\bigl[\bigl( }
  +\bigl( s_{12} \, A^{\NP}_{1234, \,\NYM} +s_{34}\, A^{\NP}_{3421, \,\NYM}\bigr)
\\& \hskip 5mm \hphantom{\times\bigl[\bigl( }
  +  \textrm{cyclic}(2,3,4)
\bigr] \,,
\label{eq:M42lneq4}
\end{aligned}
\end{equation}
where $s_{ij} = (k_i+k_j)^2$, $A_{\NeqFour}^{\rm tree}(1,2,3,4)$ is the
four-point tree-level superamplitude of $\NeqFour$ super-Yang--Mills theory and
$A^{\P}_{1234, \,\NYM}$ and $A^{\NP}_{1234, \,\NYM}$ are respectively the
color-ordered planar and nonplanar subamplitudes of pure Yang--Mills theory
with $n_+$ positive helicities and $n_-$ negative helicities, extracted from
the QCD computation in Ref.~\cite{TwoLoopQCD}.  This form holds for both
anomalous and nonanomalous superamplitudes. The latter were presented
explicitly in Ref.~\cite{N4Twoloop}. (See Section 4 of that reference for
further details.)

This simple connection between the integrated supergravity amplitudes and those
of pure Yang--Mills theory is special to the four-point amplitude.  It
follows from the $\NeqFour$ super-Yang--Mills diagram numerators'
 \textit{independence} of loop momenta.
 At higher points, we can nonetheless compute the polylogarithmic
parts of the two-loop amplitudes. A spanning set of unitarity cuts, from
which they can be determined completely, is shown in \fig{fig:spanning}.  We can
determine these terms using four-dimensional momenta
and helicities for the cut lines.  (However, one-loop amplitudes sitting 
on either side of the cut must be evaluated in $D$ dimensions.) This is 
similar to the computation of the all-multiplicity all-plus QCD amplitude in 
Ref.~\cite{Dunbarnpt}.

The following observations greatly simplify this calculation.  First, cuts that
decompose an anomalous amplitude into a product of trees automatically
vanish when the cut lines are placed in four dimensions.  This happens because
four-dimensional cuts of anomalous amplitudes at any loop order must have
an anomalous amplitude somewhere, otherwise the original amplitude
will not carry a nonzero $U(1)$ charge. Thus we need only consider those cuts
involving a one-loop anomalous amplitude, as shown in \fig{fig:spanning}b. 

\begin{figure}[tb]
  \centering
  \includegraphics[scale=0.41]{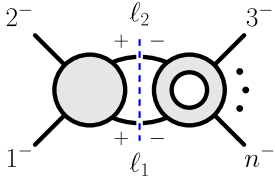}
  \caption{The only nonzero class of four-dimensional cuts of the two-loop all minus amplitude.}
  \label{fig:twoloopsinglepnptcut}
\end{figure}

Furthermore, for cuts where the configuration of external momenta is
MHV, the amplitudes on both sides of the cut must also be MHV, else
the supersums will vanish.  This limits the number of particles that
can enter the tree-level amplitude, by convention to the left of the
cut as in \fig{fig:spanning}b.  For instance, for the $M^{(0,n;2)}_{n,0}$
superamplitudes, the only suitable tree-level MHV amplitude is the
four-point one, as shown in the example of
\fig{fig:twoloopsinglepnptcut}. The other side of the cut will be an
anomalous one-loop amplitude.
%%%%%%%%%% FIGURE %%%%%%%%%%%%%
%%%%%%%%%%%%%%%%%%%%%%%%%%%%%%%
It is straightforward to compute them using \eqns{eq:treenonanom}{eq:oneloopanom},
\begin{equation}
\begin{aligned}
  \int d^4\eta_{\ell_1}d^4\eta_{\ell_2}\, & 
M^{(2,2;0)}_{n,0}(1^-,2^-,\ell_2^+,\ell_1^+) 
    \\
  &\hspace{.5cm}
  \times M^{(0,n;1)}_{n,0}(-\ell_1^-,-\ell_2^-,3^-,\ldots,n^-) \\ 
  &\hspace{-.2cm}
=  \frac12 M^{(0,n;1)}_{n,0} \frac{s_{12}^2}{(\ell_1 + k_1)^2} 
+ (\ell_1 \leftrightarrow \ell_2)\,.
  \label{eq:cutexp2}
\end{aligned}
\end{equation}
For the $M^{(1,n-1;2)}_{n,0}$ superamplitudes, there are two additional classes of
cuts: one with a four-point amplitude to the left of the cut, but where
the two external legs on the left-hand side in \fig{fig:twoloopsinglepnptcut}
are taken from different multiplet types instead of both from $\Phi^-$; the
other class, with a five-point amplitude to the left of the cut, with two
external $\Phi^-$ and one $\Phi^+$.  More generally the cuts of
$M^{(n_+,n_-;2)}_{n,0}$ can contain tree-level amplitudes with at most
$n_+ + 4$ particles.

We obtain,
\begin{align}
  M^{(0,n;2)}_{n,0}&= - M^{(0,n;1)}_{n,0}(\epsilon) 
    \sum_{i < j}^{n} s_{ij} \,
    \includegraphics[scale=0.32, trim=0 48 0 0]{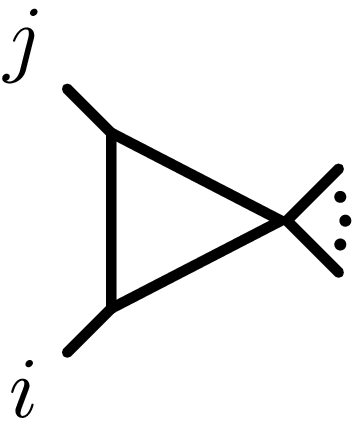}
    \,,
\vspace*{-8mm}
\nonumber
\\
M^{(1,n-1;2)}_{n,0}&= - M^{(1,n-1;1)}_{n,0}(\epsilon)  \sum_{i <  j}^{n} 
    s_{ij} \, 
  \includegraphics[scale=0.32, trim=0 48 0 0]{figures/trianglenpt}   
\label{eq:twoloopanom}
   \\
  &  \hspace{-1.5cm} - \hspace{-2pt} \sum_{i< j}^{n-1} c_{i,n;j}  
  \left( \hspace{-2pt}
    s_{ij}          \includegraphics[scale= 0.32,trim = 10 52 5 0]{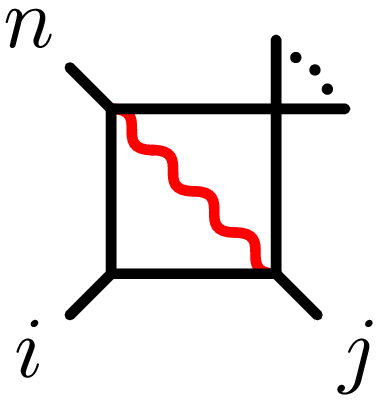} \hspace{-2pt} +    
      s_{jn}         \includegraphics[scale= 0.32,trim = 10 52 0 0]{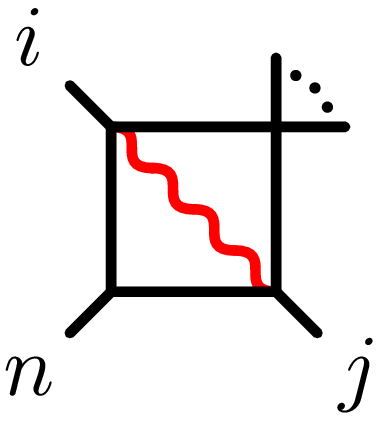} \hspace{-3pt} - \frac{2}{s_{ij}s_{in}} \includegraphics[scale=0.32, trim= 7 48 0 0]{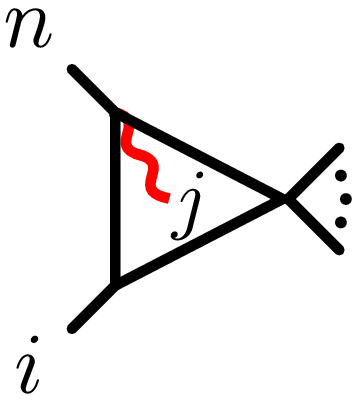} \hspace{-1pt} \right), 
\nonumber
\end{align}
up to possible additional rational terms. Here,
\begin{equation}
  c_{i,n;j} = - i \frac{\spb{i}.{n}\spa i.j^2}{\spa i.n \spa n.j^2 \;} \deltaQ\,.
\end{equation}
The would-be spurious singularities in these coefficients are in fact absent,
as they are canceled by the polylogarithmic content of the accompanying
integrals.  The integrals appearing in these expressions are the box with a
chiral numerator~\cite{ArkaniHamed2010gh},
\begin{align}
&\includegraphics[scale= 0.32,trim = 0 52 0 0]{figures/chiralboxni}  =   \int \frac{d^D\ell}{i\pi^{D/2}c_\Gamma}\frac{  \langle n |\, j \, \ell\, (\ell+i) | n]}{\ell^2(\ell+k_i)^2(\ell+k_i+k_n)^2(\ell-k_j)^2}
 \nn \\[4mm] & \hspace*{1mm}
=\Li_2\biggl(1\!-\!\frac{K^2}{s_{ij}}\biggr) 
\!+\! \Li_2\biggl(1\!-\!\frac{K^2}{s_{in}}\biggr)
\!+\! \frac12 \ln^2\bigg(\frac{s_{ij}}{s_{in}}\bigg) \!+\! \zeta_2\,;\hspace*{-6mm}
\end{align}
the normalized scalar triangle,
\begin{equation}
  \begin{aligned}
    \includegraphics[scale=0.35, trim=0 50 0 0]{figures/trianglenpt} &= \int \frac{d^D\ell}{i\pi^{D/2}c_\Gamma}\frac{ s_{ij}}{\ell^2(\ell+k_i)^2(\ell+k_i+k_j)^2} \\
  &=
   \frac{1}{\eps^2}-\frac{1}{\eps}\log(s_{ij}) 
 + \log^2(s_{ij})\,;
 \vspace{3mm}
  \end{aligned}
\end{equation}
and a chiral triangle that integrates to zero,
\begin{equation}
\includegraphics[scale=0.38, trim=0 50 0 0]{figures/trizerox} = \int \frac{d^D\ell}{i\pi^{D/2}c_\Gamma}\frac{ \langle n | \, j \, i\, \ell\,|n]^2}{\ell^2(\ell+k_i)^2(\ell+k_i+k_n)^2} 
  = 0\,.
\vspace{8pt}
\end{equation}
The latter triangle does not contribute to the amplitude but is needed to match
the cuts at the integrand level.

The results in \eqn{eq:twoloopanom} require the one-loop amplitude to higher
order in $\epsilon$.  The cross terms between the $\mathcal{O}(\epsilon)$
contributions and the leading $1/\eps$ pole in the  one-loop triangles in
\eqn{eq:twoloopanom} give rise to finite contributions.  We will call these
`IR-$\Ord(\eps)$' cross terms. Weinzierl has argued~\cite{Weinzierl2011uz} that
such contributions ultimately cancel against corresponding ones in
real-emission corrections when computing physical cross sections.
Ref.~\cite{TwoLoopQCD} provides an explicit example of such a cancellation in
QCD at two loops.

We have computed the $\mathcal{O}(\epsilon)$ terms for $n=4$ using the double
copy as explained in Refs.~\cite{CarrascoAnomaly,CurvatureSquareN4}, with the
result,
\begin{equation}
\begin{aligned}
  M_{4,0}^{(n_+,n_-;1)}(\epsilon) &= M_{4,0}^{(n_+,n_-;1)}
  \left[ 1 + \epsilon\,  g^{(n_+,n_-)}   + \mathcal{O}(\epsilon^2)\right] \,,
  \\
\end{aligned}
\label{oneloopepsilon}
\end{equation}
where,
\begin{equation}
\begin{aligned}
g^{(0,4)} &= 
\Bigl[-g_0 +  \frac{8}{6}\Bigr]
   +  \textrm{cyclic}(s,t,u) \,, \\
g^{(1,3)} &=  \Bigl[2 g_0+ \frac{1}{2}L(-s,\mu^2)
 +  \frac{8}{6}\Bigr] + \text{cyclic}(s,t,u) \,, \\ 
g_0 &= \frac{t u}{6 s^2} \bigl(L^2(t,u)+\pi ^2\bigr) 
+\frac{s^2}{3 t u} L(-s,\mu^2)\,,
\end{aligned}
\end{equation}
with $L(v,w)=\log(v/w)$, $s=s_{12}$, $t = s_{23}$ and $u=s_{13}$.  The
terms higher order in $\epsilon$ are not known explicitly beyond
$n=4$. Integrands for both the super-Yang--Mills side and a BCJ form
of the pure Yang--Mills side of the double copy for arbitrary multiplicity
were given in Ref.~\cite{BDDK,BCJOneloopYM}.  Four-dimensional
cuts do not determine these higher-order terms. They are nonetheless required by
soft limits in order to match the explicit four-point two-loop
calculation described above \eqn{eq:M42lneq4}.  There are, however,
terms in \eqn{eq:twoloopanom} for $n \ge 5$ that remain incompletely
determined: we cannot rule out additional rational terms that vanish
in all soft and complexified collinear
limits~\cite{CollinearGravity,OneLoopAnomaly} and have no overall
spurious singularities.

\Section{One-loop counterterm amplitudes}
Ref.~\cite{OneLoopAnomaly} showed that one-loop anomalous amplitudes can
be canceled by corresponding tree-level amplitudes with a single insertion of a
finite counterterm (see also Ref.~\cite{Bernard}),
\begin{equation}
\begin{aligned} 
\hskip -1mm 
\Delta S_\ct &= -\frac{1}{2(4\pi)^2}\int d^4x\, 
 \bigl(  (1-\log(1-\bar{t})) (R^+){}^2 \\
& \hskip 14mm \null 
+  (1-\log(1-t)) (R^-){\vphantom{a}}^2 \bigr)\, + \,\hbox{SUSY}\,.
\end{aligned}
\label{eq:countertermaction}
\end{equation}
In this equation, $R^\pm$ are the self-dual and anti-self-dual
components of the Riemann tensor, and $t$ and $\bar t$ are the scalars from the
two supermultiplets in \eqn{N4multiplets}.  We wish to explore this cancellation to one
additional order in perturbation theory.  Amplitudes with an insertion
of the counterterm can be obtained using the double copy for
higher-dimensional operators~\cite{BroedelDixon}. In particular we use
the double copy of $\NeqFour$ SYM with matrix elements of the $F^3$
operator added to pure Yang--Mills
theory~\cite{DixonShadmiHigherDimension}.  We can then write one-loop
four-point counterterm superamplitudes as linear combinations of
products of the former with the latter,
\begin{equation}
\begin{aligned}
  M_{\ct}^{\oneloop} (1,2,3,4) &=
  -i  c_H s_{12} s_{23} A^{\tree}_{\NeqFour} (1, 2, 3, 4) \times 
\\
& \hspace{-21mm} \Bigl(A^{\oneloop}_{F^3}(1,2,3,4)\! +\! A^{\oneloop}_{F^3}(1,3,4,2)\! +\! A^{\oneloop}_{F^3}(1,4,2,3) \Bigr) \,,
\end{aligned}
\end{equation}
where $c_H$ is an integer factor dependent on the choice of external
states (essentially on the $U(1)$ charge violation).  The
color-ordered $F^3$ amplitudes were constructed using $D$-dimensional
unitarity cuts and will be presented elsewhere~\cite{OneLoopSMEFT}.
The sum of $M_{4,0}^{(0,4;1)}$ and the corresponding $M_\ct$, or of
$M_{4,0}^{(1,3;1)}$ and its corresponding $M_\ct$, is zero up to
IR-$\Ord(\eps)$ cross terms. From the viewpoint of unitarity, adding
the finite counterterm cancels the one-loop amplitude that appears on
the right-hand side of the cut in \fig{fig:spanning}b.  The same
cancellation continues at higher points, again up to IR-$\Ord(\eps)$
cross terms.  Similar cross terms were first noted in
Refs.~\cite{Korner:1985uj,Korner:1985dt} in the context of the usual
axial anomaly in dimensional regularization. Here too they were shown
to cancel against real-emission terms in a physical quantity. The
surviving terms in our amplitudes are exactly of this type, arising
exclusively because we use a dimensional regulator.  In particular,
their presence will have no effect on ultraviolet divergences at
higher loops.

%%%%%%%%%%%% FIGURE %%%%%%%%%%%%%%
\begin{figure}[tb]
  \centering
\vskip -.3 cm 
  \begin{minipage}{0.45\columnwidth}
   \includegraphics[scale=0.60]{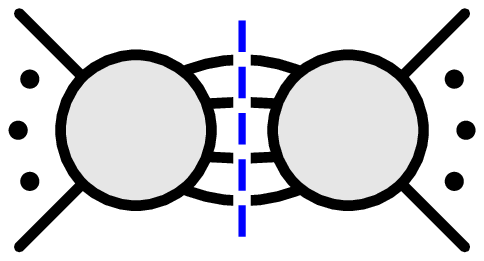}\\[-.2 cm] (a)\\[.02 cm]
   \includegraphics[scale=0.60]{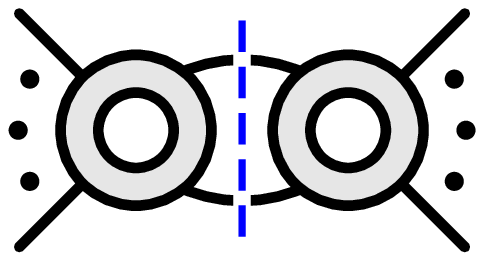}\\[-.2 cm] (c)
  \end{minipage}
%  \hfill
%\hskip 3 cm 
  \begin{minipage}{0.45\columnwidth}
  \includegraphics[scale=0.60]{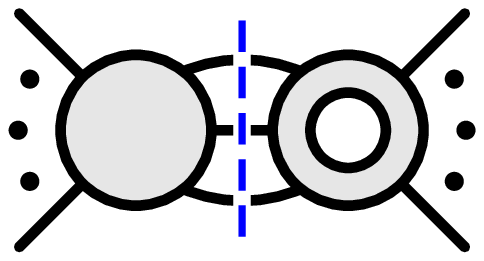}\\[-.2 cm] (b)\\[.02cm]
  \includegraphics[scale=0.60]{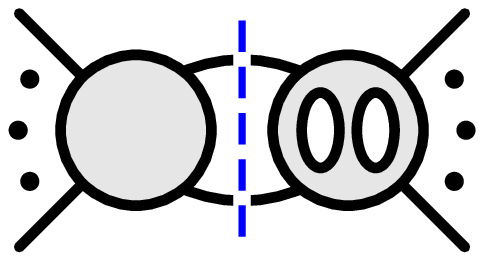}\\[-.2 cm] (d)
  \end{minipage}
  \caption{Spanning sets of unitarity cuts at three loops: (a) four-particle cuts
           (b) three-particle cuts (c) two-particle cuts and (d) additional two-particle
           cuts. Here, the annulus denotes a one-loop amplitude
           or a counterterm insertion, and the double annulus denotes a two-loop
	 amplitude. The loop amplitudes can be anomalous or not.}
  \label{ThreeLoopSpanningCuts}
\end{figure}
%%%%%%%%%%%%%%%%%%%%%%%%%%%%%%%%%%

\Section{Implications at higher loops}
Setting aside possible rational terms in the two-loop anomalous
amplitudes, all terms in the three-loop anomalous amplitudes
detectable in the four-dimensional cuts shown in
\fig{ThreeLoopSpanningCuts} will vanish in the presence of the
counterterm.  One may further conjecture that this pattern continues
to higher loops as well, making all anomalous amplitudes vanish to all
orders.

What about ultraviolet divergences?  Supersymmetry and power-counting
rule out ultraviolet divergences in $\NeqFour$ supergravity at one and
two loops. Symmetry considerations as presently understood admit a
counterterm allowing a divergence to appear at three
loops~\cite{ThreeLoopN4Counterterm, UVSymmetry}; explicit calculation
surprisingly shows that its coefficient
vanishes~\cite{ThreeLoopN4Sugra}, and the theory remains finite.  As
leading ultraviolet divergences are detectable in four-dimensional
cuts, the absence of rational terms in the two-loop amplitudes would
imply that the addition of the counterterm does not spoil the
ultraviolet finiteness of three-loop anomalous amplitudes.  A spanning
set of cuts for the three-loop four-point amplitude is shown in
\fig{ThreeLoopSpanningCuts}.  Only the four-dimensional cuts in
\fig{ThreeLoopSpanningCuts}b--d contribute to anomalous amplitudes.
The calculation of Ref.~\cite{ThreeLoopN4Sugra} shows that their sum
gives no ultraviolet divergence; the addition of the counterterm to
the theory simply adds an equal but opposite contribution, which again
vanishes.  The nonanomalous amplitudes receive new contributions when
both sides of the cuts have equal and opposite $U(1)$ charges: one
counterterm insertion and one one-loop amplitude, or two counterterm
insertions. These require further study.

At four loops, an ultraviolet divergence does appear in three distinct superamplitudes~\cite{FourLoopN4Sugra}, 
\begin{equation}
\begin{aligned}
\hskip -.2 cm 
  M^{(4)}_{\NeqFour}\Big|_{\rm div} &=
  \frac{1}{ \epsilon} \frac{(264 \zeta_3 - 1)}{288} 
  \\ &  \hspace*{-2mm}\times
  \Bigl\{M^{(2,2)}_{D^2R^4},\; \; - 3 M^{(1,3)}_{D^4t R^3},\;\; 60 M^{(0,4)}_{D^6t^2 R^2} \Bigr\} \,, \hskip .4 cm
\label{FourLoopDivergence}
\end{aligned}
\end{equation}
where $M^{(n_+,n_-)}_{\mathcal{O}}$ denotes the tree-level superamplitude with given
numbers of $\Phi^\pm$ multiplets and one insertion of the supersymmetrization of the operator $\mathcal{O}$. Its structure is quite unusual, appearing at the same order in both nonanomalous 
and anomalous four-point superamplitudes.  As the anomalous superamplitudes have
structure similar the nonanomalous amplitudes at one lower loop, one might have
expected the ultraviolet divergence to have appeared at a higher loop order
than the nonanomalous one.

The all-loop conjecture above implies that adding the counterterm
would eliminate the anomalous amplitudes from this ultraviolet
divergence.  The fate of the nonanomalous superamplitude on the
right-hand side of \eqn{FourLoopDivergence} remains to be computed
explicitly.

\Section{Acknowledgments}
We thank Radu Roiban, Enrico Herrmann, and Sergio Ferrara for useful
discussions.  Z.~B.~was supported in in part by the DOE under
Award Number DE--SC0009937.  D.~A.~K.~was supported in part by the
French Agence Nationale pour la Recherche, under grant
ANR--17--CE31--0001--01.  J.~P.-M.~is supported by the U.S.  Department of
State through a Fulbright Scholarship. J.P.-M. also thanks the Mani
L. Bhaumik Institute for generous support.

\end{document}